\documentclass[journal abbreviation, manuscript]{copernicus}

\begin{document}
\nolinenumbers

\title{Convolution Based Techniques for Computing Self Attraction and Loading in MOM6}

\Author[1,6]{Anthony}{Chen}
\Author[2,3,4]{He}{Wang}
\Author[2,5]{Brian K.}{Arbic}
\Author[1]{Robert}{Krasny}

\affil[1]{Department of Mathematics, University of Michigan, Ann Arbor, MI, USA}
\affil[2]{Department of Earth and Environmental Sciences, University of Michigan, Ann Arbor, MI, USA}
\affil[3]{UCAR, Boulder, CO, USA}
\affil[4]{NOAA-GFDL, Princeton, NJ, USA}
\affil[5]{Research School of Earth Sciences, Australian National University, Canberra, ACT, Australia}
\affil[6]{Department of Mathematics, National University of Singapore, Singapore}


\correspondence{Anthony Chen (cygnari@nus.edu.sg)}

\runningtitle{Convolution Based Self Attraction and Loading}

\runningauthor{A. Chen et al.}

\received{Feb 4, 2026}
\pubdiscuss{Apr 2, 2026} 
\revised{Oct 19, 2026}
\accepted{Nov 10, 2026}
\published{}


\firstpage{1}

\maketitle

\begin{abstract}
Self Attraction and Loading (SAL), which accounts for the deformation of the solid Earth under the load of the ocean tide and the self-gravitation of the so-deformed Earth as well as of the ocean tides themselves, is an important effect to include in numerical models of the ocean tides. Computing the SAL potential and its gradient is a challenging problem that is often accomplished using the spherical harmonics transform (SHT), but that approach has drawbacks which limit its accuracy. This work employs an alternative approach based on computing the spherical convolution integral, utilizing two implementations of the SAL Green's function, a closed form approximation, and a tabular version regularized by Ces\`aro summation. The alternative approach is accelerated by the recently developed Cubed Sphere Fast Multipole Method (CSFMM). The convolution technique is implemented in the Modular Ocean Model, version 6, where it provides more accurate tides than those obtained with the SHT when measured against altimetry-based tides. Previous investigators have cited Gibbs phenomenon as a possible source of error in SHT calculations, and these errors, along with aliasing and finite differencing errors, are avoided by the convolution. The convolution approach has different sources of error including errors in the approximate SAL Green's function, quadrature error in computing the convolution integral, and interpolation error from the CSFMM. In addition to comparing the accuracy of the  two approaches, we also present the runtime for parallel calculations using up to 1024 MPI ranks.
\end{abstract}


\introduction  

Tides are an important component of the Earth system. In many open-ocean and coastal regions, tides dominate sea level and/or current variability. Tides are also implicated as a major element in the ocean mixing cycle, which contributes significantly to the background oceanic stratification in both the coastal and open ocean \citep{simpson1974fronts,munk1998abyssal}. Tides are a result of astronomical gravitational forcing, and this tidal forcing drives periodic changes in the sea surface height (SSH) and the currents. Hence an improved understanding of tidal dynamics is necessary for developing improved models. 

Historically, tides were modeled separately from the rest of the ocean, primarily in barotropic tide models, which improved with increased computing power and data assimilation from satellite altimetry \citep{egbert1994topex,le1994spectroscopy,parke1980m2,ray2002global,schwiderski1979global,shum1997accuracy,stammer2014accuracy}. 
The first studies of tides in a baroclinic model were performed in regional configurations \citep{cummins1997simulation,kang2002two,merrifield2001generation}.
In the past two decades, as computers have improved, tides have been explicitly incorporated into ocean general circulation models \citep{arbic2022incorporating, arbic2010concurrent, muller2010effect,schiller2007explicit, thomas2001consideration, rocha2016mesoscale, waterhouse2014global}. Running such models with combined tidal and atmospheric forcing at high spatial resolution enables investigation of the interactions of barotropic and internal tides with oceanic phenomena across a wide range of scales, including mesoscale eddies, and other components of the Earth system including the cryosphere. 

An important effect to capture when modeling the tides is Self Attraction and Loading (SAL). This encompasses the physical effects of the self gravitation of the oceanic tidal elevations, the elastic response of the Earth's crust to the loading of the ocean tide, and resulting changes in the Earth's gravitational potential due to the gravitation of the deformed solid Earth \citep{hendershott1972effects}. SAL has a large effect on tides, changing their amplitude by up to 20\% and impacting their phase significantly as well \citep{gordeev1977effects}. 
A simple computationally efficient approach to computing SAL uses the so-called scalar approximation \citep{accad1978solution}, in which the SAL forcing term is approximated as a constant multiplied by the tidal elevation.

However, the scalar approximation is less accurate than a treatment using spherical harmonics, which properly account for the response of the ocean and solid Earth to harmonics of different scales \citep{ray1998ocean}. 
Many earlier studies that included the spherical harmonics SAL in tide models took an iterative approach using offline computations \citep{arbic2004accuracy,egbert2004}, where the SAL was calculated outside of the model run. The iterative approach is cumbersome. Early attempts to use the spherical harmonics SAL inline, as the model is running, found it to be too computationally expensive \citep{stepanov2004parameterization}. This is because calculating SAL requires global communication, in contrast to other terms in the momentum equations of numerical ocean models. 
More recent modeling using an inline spherical harmonics SAL has been aided by improvements in computational power \citep{barton2022global,brus2023scalable,schindelegger2018can,wang2024improving}, 
yet this approach still encounters limitations on accuracy
around coastlines (more below), and the fact that spherical harmonics, being global quantities, 
cannot easily be used in regional models. 

Another approach to computing SAL, often used in the geodesy community, 
is based on a convolution of the sea surface height with a Green's function \citep{farrell1972deformation,schrama2005three,agnew2012spotl,martens2019loaddef}. This approach has been explored in post-processing contexts, and past efforts at incorporating this method into an ocean model have faced difficulties with the expense of computing the convolution~\citep{stepanov2004parameterization}.
Here we employ the recently developed Cubed Sphere Fast Multipole Method (CSFMM; \cite{chen2025cubed})
to reduce the cost of computing the integral
and
enable efficient inline SAL calculations.
The convolution SAL is implemented in the
Modular Ocean Model, version 6,
and
the results show that it provides more accurate tides
than those obtained with the spherical harmonics approach.

The paper is organized as follows. Section~\ref{sec:methods} 
presents the governing equations,
describes the various methods for computing SAL,
introduces our convolution SAL technique, 
and presents results comparing the different methods using a snapshot of the tidal sea surface height. 
Section~\ref{sec:simulation} describes the model configuration used to test the convolution SAL. Section~\ref{sec:results} presents 
results comparing tidal errors and computational efficiency using the convolution and spherical harmonics approaches. 
Conclusions are presented in Section~\ref{sec:conclusions}.

\section{Methods and Implementation}
\label{sec:methods}

The Modular Ocean Model version 6 (MOM6)
is a widely used open source numerical ocean general circulation model \citep{adcroft2019gfdl}. The model is hydrostatic and solves either a Boussinesq or non-Boussinesq equation set.
Here we use a single layer configuration 
in which the dynamical equations reduce to the shallow water equations on a sphere,
\begin{align}
    \frac{\partial\vec{u}}{\partial t}+(\vec{u}\cdot\nabla)\vec{u}+\vec{f}\times\vec{u}&=-g\nabla \eta+\mathcal{F}\label{eq:ocean-momentum},\\
    \frac{\partial \eta}{\partial t}+\nabla\cdot(h\vec{u})&=0,\label{eq:ocean-mass-conservation}
\end{align}
where $t$ is time, $\vec{u}$ is the horizontal fluid velocity, $\vec{f}=f\vec{e}_r$, where $f$ is the Coriolis parameter and $\vec{e}_r$ is the radial outward unit vector, $\eta$ is the perturbation sea surface height, $g$ is acceleration due to gravity, $\mathcal{F}$ are other forces (for instance, friction and tidal forcing terms), and $h$ is the total water column thickness. If the ocean floor has height $b(\vec{x})$, 
then $\eta(t,\vec{x})=h(t,\vec{x})+b(\vec{x})$. Equation~\ref{eq:ocean-momentum} expresses the balance of momentum, and Eq.~\ref{eq:ocean-mass-conservation} expresses the conservation of mass.
For multi-layer configurations, 
this system can be supplemented with thermodynamic equations,
and in that case it is often replaced by the more general Boussinesq or non-Boussinesq equations. 
The horizontal grid is based on a global tripolar discretization with an Arakawa C-grid. 
In this work 
the forcing term has several components,
\begin{equation}\label{eq:forcing}   \mathcal{F}=g\nabla(\eta_{\mathrm{EQ}}+\eta_{\mathrm{SAL}})+\vec{F}_{\mathrm{wave}}+\vec{F}_{\mathrm{quad}}+\vec{F}_{h},
\end{equation}
where 
$\eta_{\mathrm{EQ}}$ represents equilibrium tidal forcing corrected for the effects of solid earth body tides \citep{hendershott1972effects}, 
$\eta_{\mathrm{SAL}}$ is the SAL potential, $\vec{F}_{\mathrm{wave}}$ is 
the linear wave drag, 
$\vec{F}_{\mathrm{quad}}$ is quadratic bottom 
boundary layer drag, and $\vec{F}_h$ is horizontal viscosity. 
The linear wave drag 
includes a tunable parameter $\chi$.
Further details on the wave drag scheme are given by 
\cite{jayne2001parameterizing} and \cite{buijsman2015optimizing}. We refer to $\eta_{\mathrm{SAL}}$ as the SAL potential and $g\nabla\eta_{\mathrm{SAL}}$ as the SAL acceleration. 

\subsection{Spherical Harmonics for Computing SAL}

Recent efforts to compute SAL inline \citep{schindelegger2018can,barton2022global,brus2023scalable,wang2024improving} have used spherical harmonics. 
In this technique, the spherical harmonics decomposition of the sea surface height $\eta$ (i.e., the tidal mass anomaly) is computed, with 
spherical harmonic coefficients $\widehat{\eta}_{n,m}$
given by
\begin{equation}    \widehat{\eta}_{n,m}=\int_{0}^{2\pi}\int_0^{\pi}Y_n^m(\theta,\phi)\eta(\theta,\phi)\cos\theta d\theta d\phi,
\label{eq:eta_coefficients}
\end{equation}
where $Y_n^m$ is the degree $n$ order $m$ spherical harmonic, $\theta$ is the latitude, and $\phi$ is the longitude.
In practice two approaches are available to compute these coefficients. 
One approach uses a 
fast spherical harmonics transform
to reduce the operation count \citep{barton2022global}, 
but this approach faces challenges with parallel scaling,
and 
additional interpolation steps are needed
if the sea surface height is not given on a latitude-longitude grid. 
Another approach computes the integrals
in Eq.~\eqref{eq:eta_coefficients}
directly without using a fast transform
as in~\cite{brus2023scalable} and~\cite{wang2024improving}. 
In this case the operation count is not reduced,
but the direct calculation is better suited for parallelization.

Given the coefficients $\widehat{\eta}_{n,m}$,
the SAL potential has the expansion 
\begin{equation}\label{eq:sal-sh}
\eta_{\mathrm{SAL}}(\theta,\phi)=\sum_{n=0}^{\infty}\sum_{m=-n}^n\frac{3\rho_w(1+k_n'-h_n')}{\rho_e(2n+1)}\widehat{\eta}_{n,m}Y_n^m(\theta,\phi),
\end{equation}
where  
$\rho_w$ and $\rho_e$ are the average density 
of seawater and the Earth in kilograms per cubic meter, 
and $k_n'$ and $h_n'$ are the modified load Love numbers (LLNs) as discussed by \cite{farrell1972deformation}. 
We use average values for the real Earth,
$\rho_w=1035\;\unit{kg\,m^{-3}}$, and
$\rho_e=5517\;\unit{kg\,m^{-3}}$. 
When a multi-layer model is used, the sea surface height is replaced with the bottom pressure anomaly
and the 
coefficients~$\widehat{\eta}_{n,m}$
are adjusted appropriately. 
In practice the sum in Eq.~\eqref{eq:sal-sh}
is truncated, 
for example going up to degree $N=40$ \citep{brus2023scalable}. Numerical experiments have found that using larger degrees 
has not significantly reduced the tidal error. 
The fact that tidal errors do not decrease past $N=40$ can be attributed to Gibbs phenomenon and aliasing errors that
occur when high wavenumber components in the 
solution are projected to low wavenumbers 
if the underlying spatial grid is not 
sufficiently fine~\citep{sanso1990aliasing,jekeli1996spherical}. These errors will be largest in coastal regions, and coastal tides have a significant back effect on the open ocean tides~\citep{arbic2007resonance,arbic2009tidal,arbic2010coupled}. 
After computing the SAL potential $\eta_{\mathrm{SAL}}$, the SAL acceleration $g\nabla\eta_{\mathrm{SAL}}$ 
is computed by finite differences,
which can lead to additional errors
by amplifying
high wavenumber errors in $\eta_{\rm SAL}$.

\subsection{Gibbs Phenomenon}

The Gibbs phenomenon  
occurs when truncating the Fourier series of a 
periodic function $f$ with a jump discontinuity \citep{gottlieb1997gibbs}. 
If $f$ has a jump discontinuity, then finite truncations of the Fourier series of $f$ will overshoot and undershoot around the discontinuity. This overshoot has a magnitude of approximately $9\%$, and this oscillatory error does not diminish 
as more terms are used, 
but instead it becomes sharper and more concentrated around the discontinuity. 
In the context of computing SAL, 
the Gibbs phenomenon may arise from the finite truncation of a spherical harmonics series.
When computing the spherical harmonic coefficients of the sea surface height $\widehat{\eta}_{m,n}$, 
the data field $\eta$ is set to zero over land.
While physically reasonable, 
this essentially introduces a jump discontinuity along coastlines, which adversely affects the convergence of 
the series. 
The presence of the Gibbs phenomenon in spherical harmonics calculations is a longstanding problem~\citep{weyl1910gibbs,gelb1997resolution}
and the geodesy community has advocated for an alternative
based on convolution with a Green's function~\citep{merriam1980series,schrama2005three}. 

\subsection{Ces\`aro Summation}

Ces\`aro summation is a general method for 
improving the convergence properties of a series.
In the case of the spherical harmonics series in 
Eq.~\eqref{eq:eta_coefficients},
one defines the partial sums 
\begin{equation}
    S_J=\sum_{n=0}^J\sum_{m=-n}^n\frac{3\rho_w(1+k_n'-h_n')}{\rho_e(2n+1)}\widehat{\eta}_{n,m}Y_n^m(\theta,\phi),
\end{equation}
and then averages them to obtain the approximation
\begin{equation}
    \eta_{\mathrm{SAL}}(\theta,\phi)\approx\frac{1}{N+1}\sum_{J=0}^NS_J,
\end{equation}
which is equivalent to modifying the coefficients
of the original series,
\begin{equation}
    \eta_{\mathrm{SAL}}(\theta,\phi)\approx\sum_{n=0}^N\sum_{m=-n}^n\left(1-\frac{n}{N+1}\right)\frac{3\rho_w(1+k_n'-h_n')}{\rho_e(2n+1)}\widehat{\eta}_{n,m}Y_n^m(\theta,\phi).
\end{equation}
This technique can be implemented by modifying the existing spherical harmonics computation with no additional cost. 
Ideally, this technique will reduce the Gibbs phenomenon along coastlines, 
however at the same time,
it introduces an error by modifying the LLNs. 
As far as we know,
Ces\`aro summation has not previously been 
applied in
spherical harmonics SAL computations. 

\subsection{Convolutions for Computing SAL}

It is well known that the SAL potential 
can be computed as a convolution on the unit sphere $S$,
\begin{equation}\label{eq:sal-conv}
    \eta_{\mathrm{SAL}}(\vec{x})=\int_SG_{\mathrm{SAL}}(\vec{x}\cdot\vec{y})\eta(\vec{y})dS(\vec{y}),
\quad \vec{x}, \vec{y} \in S,
\end{equation}
where the SAL Green's function is 
\begin{equation}\label{eq:salgf}
    G_{\mathrm{SAL}}(\vec{x},\vec{y})=\sum_{n=0}^{\infty}\frac{3\rho_w(1+k_n'-h_n')}{4\pi\rho_e}P_n(\vec{x}\cdot\vec{y}),
\end{equation}
and $P_n$ is the $n$-th degree Legendre polynomial  \citep{farrell1972deformation,farrell1973discussion}. 
Given a set of grid cells covering the sphere,
the integral in Eq.~\eqref{eq:sal-conv}
is computed
using a numerical quadrature scheme.
However, in order to use this Green's function, we cannot compute a truncated Legendre series in every single pairwise interaction in the discretized $N$-body sum, as this would be prohibitive in an ocean model where the SAL term 
is computed inline at every time step. There are two possible approaches that we can take. The first is to numerically sum the Green's function to a high degree outside of the model and tabulate the results, interpolating between values as needed. This approach has been taken previously~\citep{farrell1972deformation,guo2004green,schrama2005three,martens2019loaddef}. The other is to approximate the LLNs in order to find a closed form Green's function. 
Working with the LLNs in \cite{wang2012load}, 
we empirically determined the following approximations, 
\begin{equation}
k_n'\approx a_1/n,\quad h_n'\approx b_0+b_1/n,
\quad a_1 = -2.7, 
\quad b_0 = -6.21196, 
\quad b_1 = 6.1.
\end{equation}
Then substituting in Eq.~\eqref{eq:salgf}
and using the identities \citep{gradshteyn2014table},
\begin{equation}
    \sum_{n=0}^{\infty}P_n(x)=\frac{1}{\sqrt{2(1-x)}},\quad \sum_{n=1}^{\infty}\frac{1}{n}P_n(x)=-\ln\left(\sqrt{\frac{1-x}{2}}+\frac{1-x}{2}\right),
\end{equation}
a closed form approximation for the Green's function
is obtained,
\begin{equation}\label{eq:salgf2}
    G_{\mathrm{SAL}}(\vec{x},\vec{y})\approx\frac{3\rho_w}{4\pi\rho_e}\left(\frac{1-b_0}{\sqrt{2-2\vec{x}\cdot\vec{y}}}-(a_1-b_1)\ln\left(\sqrt{\frac{1-\vec{x}\cdot\vec{y}}{2}}+\frac{1-\vec{x}\cdot\vec{y}}{2}\right)\right).
\end{equation}
From Eq.~\eqref{eq:forcing} 
we see that what is needed is the gradient of the SAL potential,
$\nabla\eta_{\rm SAL}$.
Assuming $\eta_{\rm SAL}$ has been computed,
the gradient could be obtained by finite differences,
but this could lead to large errors in coastal regions.
Instead, taking the gradient of Eq.~\eqref{eq:sal-conv}, one obtains
\begin{equation}\label{eq:salconvgrad}
    \nabla\eta_{\mathrm{SAL}}(\vec{x})=\int_S\nabla G_{\mathrm{SAL}}(\vec{x} \cdot \vec{y})\eta(\vec{y})dS(\vec{y}),
\end{equation}
where the gradient falls on the $\vec{x}$-variable of the Green's function. 
The chain rule yields
the gradient of the Green's function,
\begin{equation}
    \nabla G_{\mathrm{SAL}}(\vec{x}\cdot\vec{y})=\frac{dG_{\mathrm{SAL}}(\cos\gamma)}{d\cos\gamma}\nabla\cos\gamma(\vec{x},\vec{y}),
\end{equation} 
where $\vec{x}\cdot\vec{y} = 
\cos\gamma(\vec{x},\vec{y})$
and
$\gamma(\vec{x},\vec{y})$ is the great circle angle between $\vec{x}$ and $\vec{y}$. 
Then taking these derivatives yields
\begin{equation}\label{eq:salconvker}
    \nabla G_{\mathrm{SAL}}(\vec{x},\vec{y})=\frac{3\rho_w}{4\pi\rho_eR_E\sqrt{1-x_3^2}}\begin{bmatrix}
        y_3(1-x_3^2)-x_3(x_1y_1+x_2y_2)\\x_1y_2-x_2y_1
    \end{bmatrix}\left(\frac{1-b_0}{(2-2\vec{x}\cdot\vec{y})^{3/2}}-\frac{(a_1-b_1)(2\vec{x}\cdot\vec{y}+\sqrt{2-2\vec{x}\cdot\vec{y}}}{2((\vec{x}\cdot\vec{y})^2-1)}\right),
\end{equation}
where $(x_1,x_2,x_3)$ and $(y_1,y_2,y_3)$ 
are the Cartesian coordinates of $\vec{x}$ and $\vec{y}$. The convolution using this closed form Green's function will be referred to later on as the closed form convolution. 

To use a tabular Green's function, we once again take the gradient of the Green's function, written as 
\begin{equation}
    \nabla G_{\mathrm{SAL}}(\vec{x}\cdot\vec{y})=\nabla\cos\gamma(\vec{x},\vec{y})\sum_{n=0}^{\infty}\frac{3\rho_w(1+k_n'-h_n')}{4\pi\rho_e}P_n'(\vec{x}\cdot\vec{y}).
\end{equation}
The series is truncated at $N=46342$, which is the number of LLNs provided by~\cite{wang2012load}. The resulting finite sum is evaluated for $\vec{x}\cdot\vec{y}$ on the interval $[-1,1]$, with a grid spacing of $\Delta x=0.001$, for a total of $2001$ evaluations. For $\vec{x}\cdot\vec{y}=\pm1$, the sum is divergent, as $P_n'(1)=n(n+1)/2$ and $P_n'(-1)=(-1)^{n+1}n(n+1)/2$, and $1+k_n'-h_n'$ does not decay to $0$. Additionally, in the interval $(-1,1)$, the sum is highly oscillatory and displays poor convergence. To remedy this, Ces\`aro summation is applied. At intermediate values where the sum was not computed, linear interpolation is used to find an approximate value. The convolution using this Ces\`aro regularized tabular Green's function will be referred to later on as the Ces\`aro tabular convolution. Several other techniques for computing this sum and regularizing the oscillations exist, including the use of Kummer's transformation as in~\cite{guo2004green} and~\cite{martens2019loaddef}. We also tested these techniques, but the resulting tide simulations perform worse than the Ces\`aro tabular convolution, which we will focus on. Even though the LLNs are exact from $N=1$ to $46342$, this Green's function is still an approximation, as the Legendre series is truncated and the singularity at $\mathbf{x}\cdot\mathbf{y}=1$ is not represented in a table. For both forms of the Green's function, tuning is possible by multiplying through by a scalar parameter. This can be used to match the acceleration magnitude with the acceleration computed using the SHT SAL. 

\subsection{Fast Summation for Accelerating Convolutions}

We employ a midpoint discretization of Eq.~\eqref{eq:salconvgrad}
yielding an approximation for the gradient of the SAL potential,
\begin{equation}
\nabla\eta_{\mathrm{SAL}}(\vec{x}_i) \approx 
\sum_{j=1,j\neq i}^{N_0}\nabla G_{\mathrm{SAL}}(\vec{x}_i,\vec{x}_j)\eta(\vec{x}_j)A_j,\quad i=1, \ldots, N_0,
\label{eq:N-body-sum}
\end{equation}
where $N_0$ is the number of grid cells
used in MOM6,
$\vec{x}_i$ is the midpoint of a grid cell, 
and  
$A_i$ is the grid cell area. 
As the SAL acceleration is computed at cell centers, interpolation is performed to obtain values 
at cell edges. 
The kernel singularity at $i=j$ is skipped,
which is consistent with taking the
Cauchy principal value of the integral.
If the sums in Eq.~\eqref{eq:N-body-sum}
are computed directly using loops over the indices 
$i$ and $j$,
the operation count scales like $O(N_0^2)$, 
which is prohibitive for reaching high resolution. 
Various fast summation methods have been developed
to accelerate the computation of such sums
including the treecode \citep{barnes1986hierarchical}
and
Fast Multipole Method \citep{greengard1987fast}.

Here we employ the recently developed
Cubed Sphere Fast Multipole Method (CSFMM) 
to reduce the operation count to $O(N_0)$~\citep{chen2025cubed}. 
The CSFMM applies barycentric Lagrange interpolation 
to approximate the kernel \citep{berrut2004barycentric},
with proxy points defined by 
a tensor product grid of Chebyshev points 
in a quadtree of cubed sphere grid cells.
The scheme introduces an error whose
amplitude is controlled by two user-specified parameters:
the interpolation degree $n_0$ of the barycentric Lagrange interpolation used in the approximations of the kernel, 
and
a parameter $\theta_0$ controlling the size and
separation of the cubed sphere grid cells. 
In contrast to spherical harmonics approximations,
the interpolation is local,
and
this approach does not encounter the Gibbs phenomenon.
In addition, the quadtree structure of the 
cubed sphere grid cells
naturally adapts to the irregular distribution of ocean grid points around coastlines. A detailed discussion of the error behavior of spherical convolutions and the CSFMM is presented in~\cite{chen2025cubed}. 

\subsection{Comparison of Methods}

\begin{figure}
    \centering
    \includegraphics[width=0.9\linewidth]{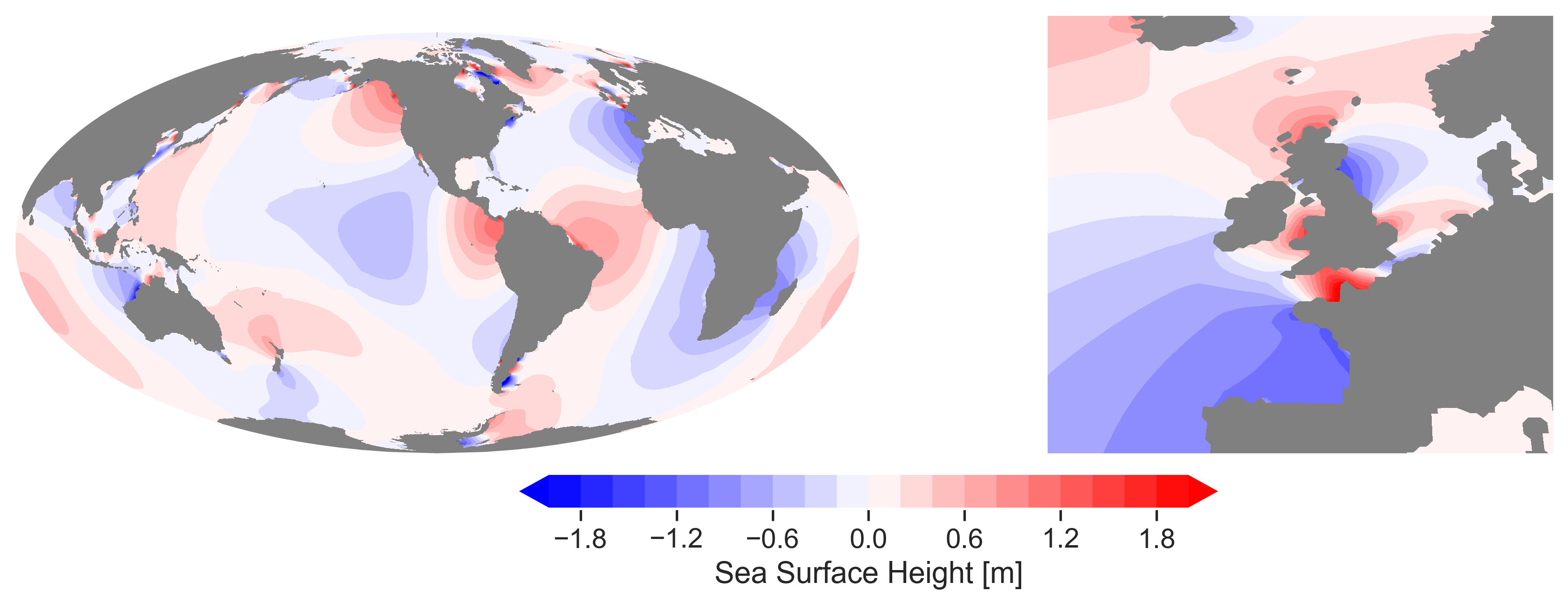}
    \caption{Global snapshot of tidal sea surface height used to compare different SAL computation methods. The right panel is a zoomed version of the left panel focusing on the Northeast Atlantic, a region of exceptionally large coastal tides. The SSH snapshot is on a MOM6 grid with $0.08^{\circ}$ nominal grid spacing, taken from an existing MOM6 barotropic tidal simulation.}
    \label{fig:sshprofile}
\end{figure}

\begin{figure}
    \centering
    \includegraphics[width=0.9\linewidth]{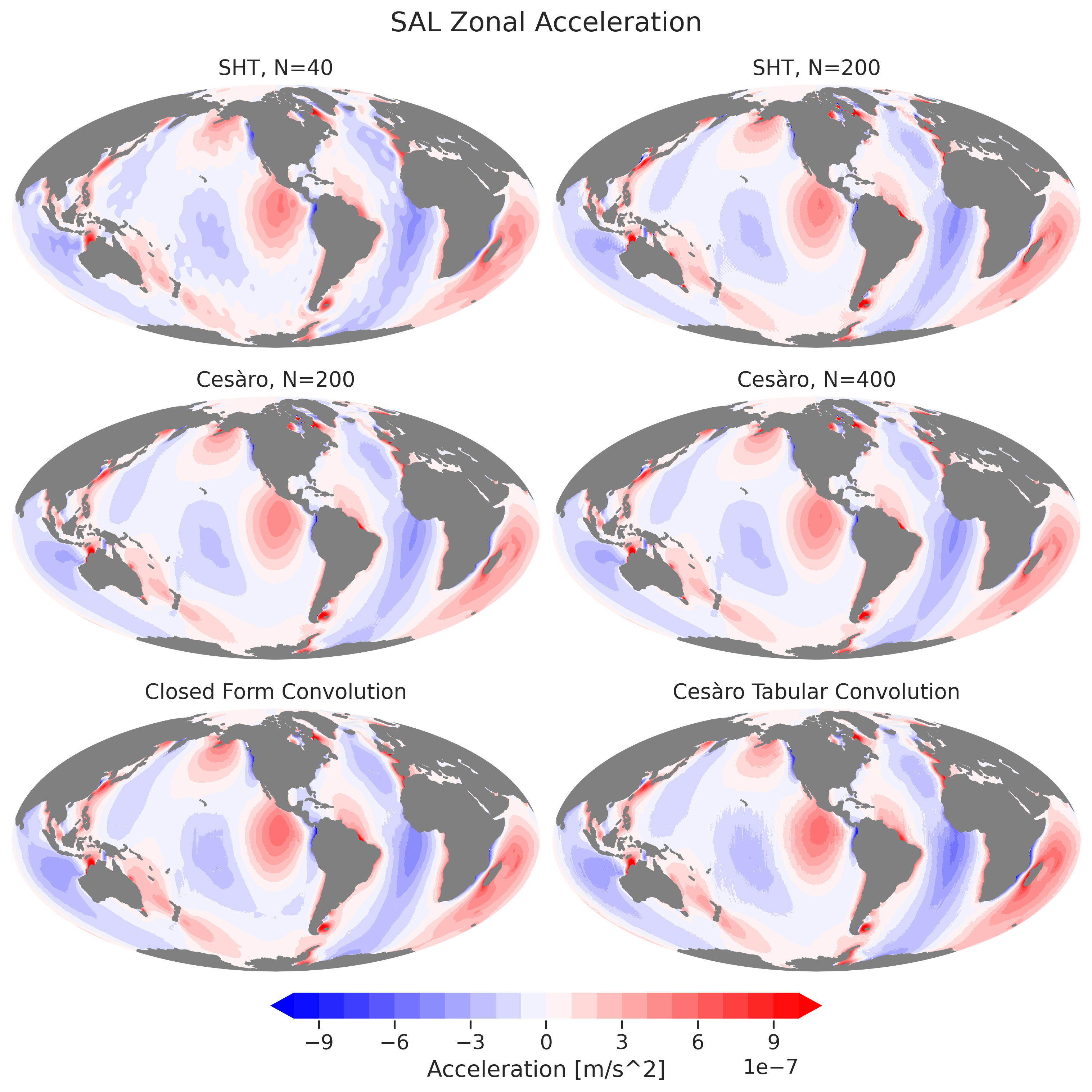}
    \caption{SAL zonal acceleration for the tidal sea surface height snapshot in Fig.~\ref{fig:sshprofile}, computed using spherical harmonics transform (SHT) with $N=40, N=200$, Ces\`aro regularized SHT with $N=200, N=400$, and both convolution approaches.}
    \label{fig:salzonalaccels}
\end{figure}

\begin{figure}
    \centering
    \includegraphics[width=0.9\linewidth]{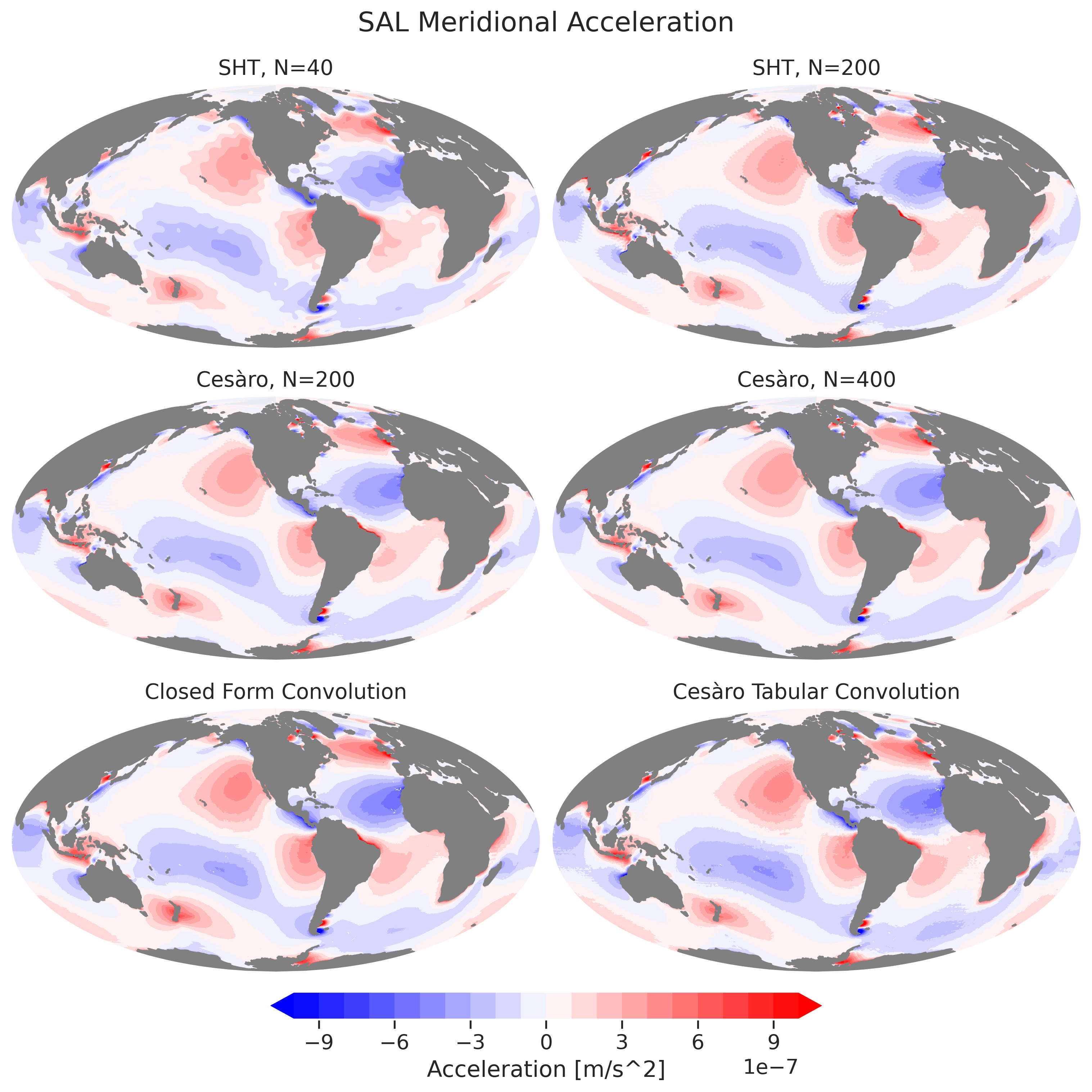}
    \caption{SAL meridional acceleration for the sea surface height snapshot in Fig.~\ref{fig:sshprofile}, computed using spherical harmonics transform (SHT) with $N=40, N=200$, Ces\`aro regularized SHT with $N=200, N=400$, and both convolution approaches. }
    \label{fig:salmeridaccels}
\end{figure}

\begin{figure}
    \centering
    \includegraphics[width=0.9\linewidth]{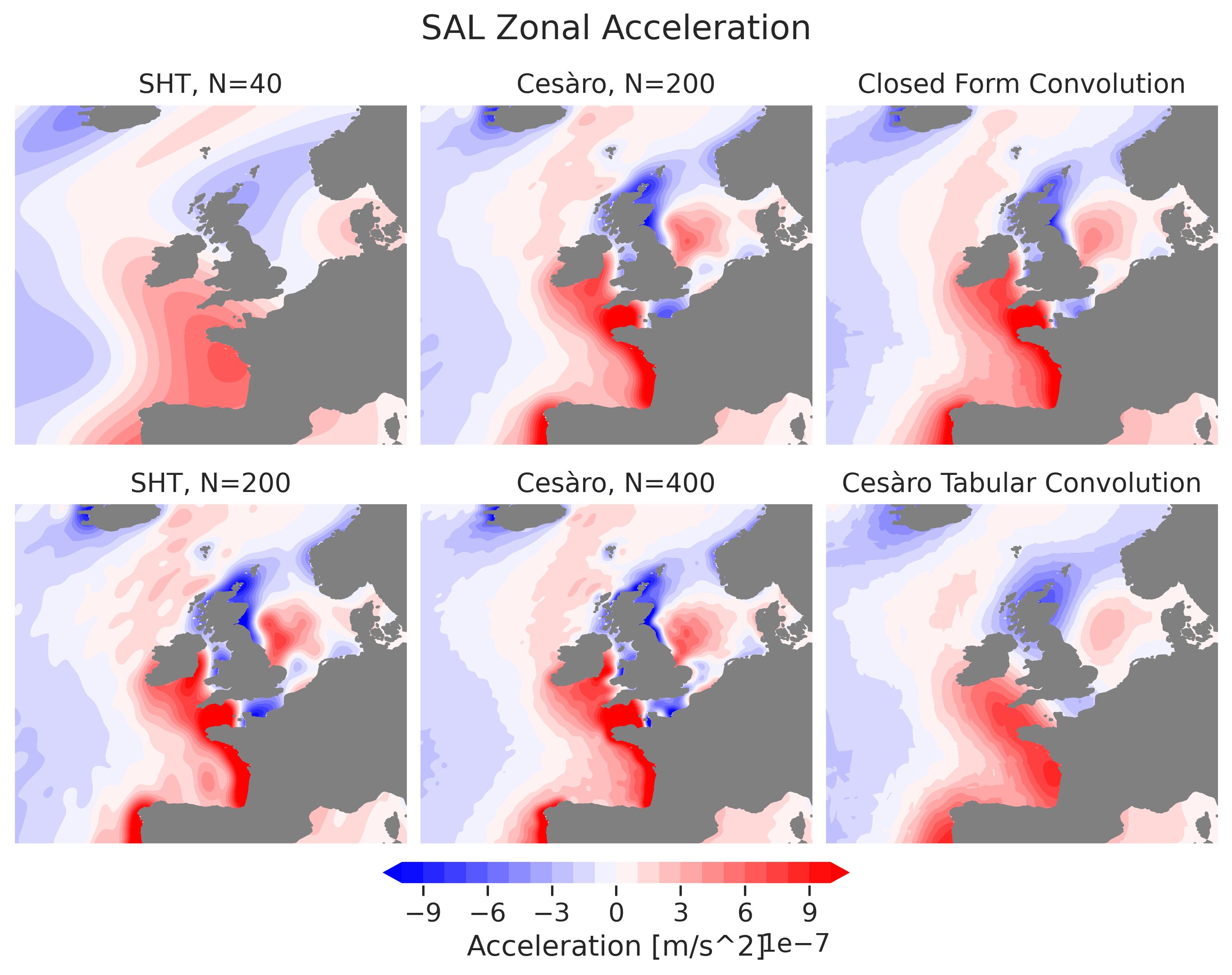}
    \caption{As in Fig.~\ref{fig:salzonalaccels}, focusing on the Northeast Atlantic.}
    \label{fig:sallocalzonalaccels}
\end{figure}

\begin{figure}
    \centering
    \includegraphics[width=0.9\linewidth]{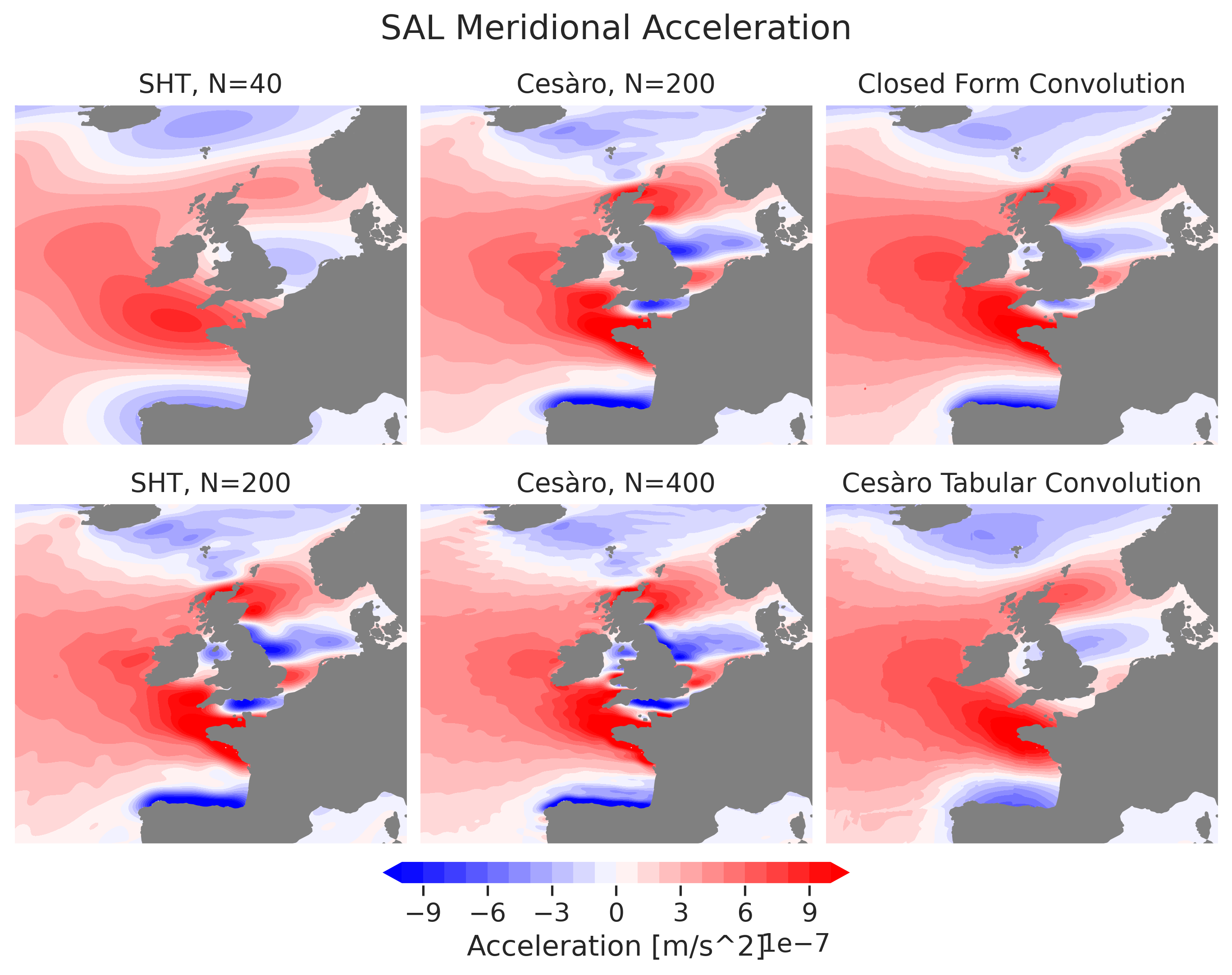}
    \caption{As in Fig.~\ref{fig:salmeridaccels}, focusing on the Northeast Atlantic.}
    \label{fig:sallocalmeridaccels}
\end{figure}

\begin{figure}
    \centering
    \includegraphics[width=0.7\linewidth]{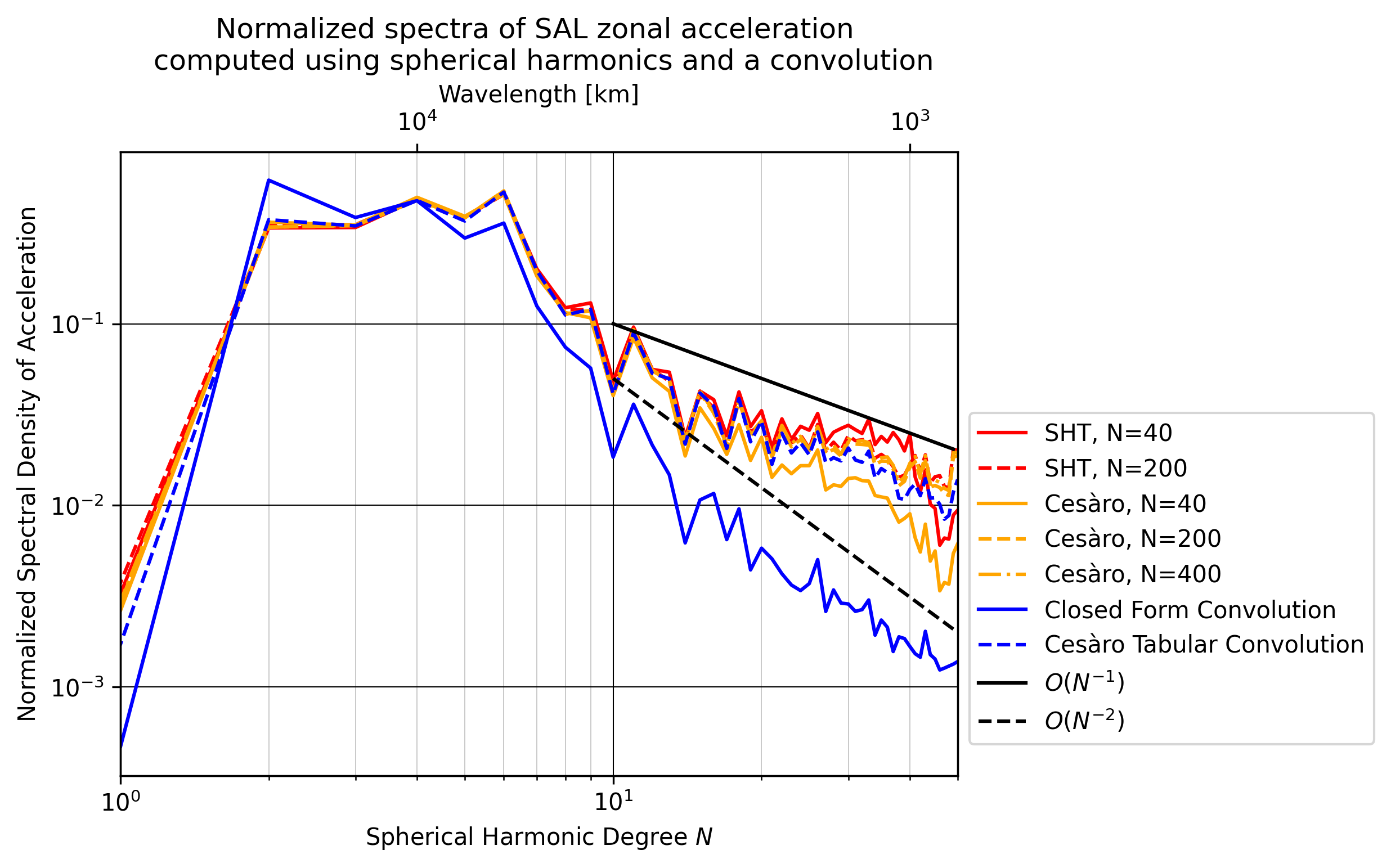}
    \caption{Normalized spectral density of the global SAL zonal acceleration field shown in Fig.~\ref{fig:salzonalaccels}. 
    A spherical harmonic of degree $N$
    has wavelength $2\pi R_E/N$. }
    \label{fig:salaccelspec}
\end{figure}

Before discussing the implementation of convolution SAL in MOM6,
we compare the various methods for computing SAL without any confounding non-SAL factors. 
We compute the SAL acceleration on a MOM6 grid with $0.08^{\circ}$ nominal grid spacing 
for the global snapshot of
tidal sea surface height (SSH) shown in Fig.~\ref{fig:sshprofile} using the various methods described in the previous sections. The SSH snapshot comes from a tidal MOM6 run, with SAL computed using the SHT with $N=40$. 
In the following tests,
the SHT is used with 
maximum degree $N=40$ and $N=200$, 
and the Ces\`aro SHT is used with 
$N=200$, and $N=400$. 
These are compared with the two convolution approaches, using the closed form Green's function and the Ces\`aro tabular Green's function. The CSFMM is used to accelerate the convolution, with CSFMM parameters $n_0=2$ and $\theta_0=0.9$. 
Global zonal and meridional accelerations are 
plotted in Figs.~\ref{fig:salzonalaccels} and~\ref{fig:salmeridaccels}, and 
accelerations focused on the Northeast Atlantic
are plotted in Figs.~\ref{fig:sallocalzonalaccels} and~\ref{fig:sallocalmeridaccels}.
Figure~\ref{fig:salaccelspec}
plots the spectral density of the zonal acceleration in Fig.~\ref{fig:salzonalaccels}. As all the calculations are based on the same reference SSH snapshot, differences can be attributed to the different methods of computing SAL, as opposed to confounding effects from non-SAL factors in the model. 

Looking first at the global zonal and meridional accelerations, 
there is general agreement among all the results.
As the maximum degree increases from 
$N=40$ to $N=200$ in the SHT computation, 
one observes less large scale waviness in the contour lines, due to a reduction in truncation error. 
While differences between the $N=40$ and 
$N=200$ results may appear large, 
they are actually quite small, less than $10^{-8}\;\unit{m\,s^{-2}}$ in the open ocean. Additionally, the large scale patterns of the Ces\`aro SHT result broadly match the unmodified SHT result, with small differences. 
This is because Ces\`aro summation
has the largest effect on small scales at high wavenumber, 
while the plots mainly depict the low wavenumber large scale structure. Lastly, there are some differences in the large scale structure between the closed form convolution result and the four SHT computations. This is because the LLN approximations used in deriving the closed form SAL Green's function are least accurate at low wavenumber and have the greatest impact at large scales. 

Next consider the regional accelerations in Figs.~\ref{fig:sallocalzonalaccels} and~\ref{fig:sallocalmeridaccels}. 
Clear differences can be seen between the SHT results with $N=40$ and $N=200$,
where $N=200$ resolves small scale features in coastal regions
while predicting larger coastal amplitudes.
As coastal regions
are spanned by relatively few grid cells, aliasing of high degree spherical harmonics
may contribute to these large amplitudes. 
Comparing the 
$N=200$ SHT result with the 
$N=200$ Ces\`aro SHT result, 
one sees the regularization effect of the latter
as maximum amplitudes are smaller and the waviness in the contours is reduced.
The $N=400$ Ces\`aro result
further reduces the waviness and resolves small scale features, but also increases the coastal amplitudes. 
The convolution results are similar to the 
high degree SHT results, 
but it feature smaller coastal amplitudes. The LLN approximations used in deriving the closed form SAL Green's function are most accurate at high wavenumber and small scales. Additionally, at this level of detail, we see that the Ces\`aro tabular convolution has some differences from the closed form convolution. While both feature smaller coastal amplitudes than the calculations with spherical harmonics, the Ces\`aro tabular convolution also has some waviness in the contour lines that is not present with the closed form. 
While the convolution approach has
errors in the approximate SAL Green's function,
quadrature error in computing the integral,
and
interpolation error from the CSFMM,
it avoids the Gibbs phenomenon, 
aliasing error, and finite difference errors 
present in the SHT calculations. 

To gain further insight into the methods,
Fig.~\ref{fig:salaccelspec}
shows the normalized spectral density of the
SAL zonal acceleration.
There is general agreement among the four SHT results, which decay roughly as $O(N^{-1})$. 
Comparing the $N=40$ SHT with the 
$N=40$ Ces\`aro SHT, 
the latter spectrum falls off slightly faster at high wavenumber. 
A similar trend is seen at $N=200$,
but to a lesser degree. 
In contrast, the closed form convolution spectrum
falls off significantly faster at high wavenumber, decaying like $O(N^{-2})$,
which is consistent with the smoother 
acceleration fields
seen in Fig.~\ref{fig:sallocalzonalaccels} and Fig.~\ref{fig:sallocalmeridaccels}. However, the closed form convolution has a significant mismatch in the degree-$1$ spherical harmonic component, which relates to tidal geocenter motion~\citep{desai2014consideration}. With the closed form convolution, it is likely that the approximations made in the derivation of the closed form Green's function cause this discrepancy. Additionally, with the Ces\`aro tabular convolution, the discrepancy at degree-$1$ is much smaller. However, we also observe that the tabular convolution has a spectra that decays in line with the spectra of the SHT calculations. This is consistent with the wavier contour lines seen in the acceleration maps. 

\section{Simulation Details}
\label{sec:simulation}

The convolution/CSFMM technique described above
was implemented in MOM6.
The model was run with zero initial velocity
and 
M$_2$ tidal forcing
for 20 days with time step $\Delta t=180\;\unit{s}$.
The SAL acceleration was computed inline at every time step
and
the last three days of output are used to evaluate the results. 
A single level configuration was used with 
two different grids,
one with nominal spacing $0.36^{\circ}$ 
(typical of long duration low-resolution climate simulations), 
and one with nominal spacing $0.08^{\circ}$ 
(typical of short duration high-resolution simulations). 
The model was run on the Derecho computer system at NCAR~\citep{derecho} and Vanda at NUS. 
The topography and wave drag are based on GEBCO 2023.
Several configurations are tested:
SHT with $N=40, 200$, 
Ces\`aro SHT with $N=40, 200, 400$, 
and the convolution, using both the closed form and Ces\`aro tabular Green's functions. 

\subsection{Measuring Error}

To measure the error in the model tides, the model output is compared with the TPXO9 dataset \citep{erofeeva2018tpxo9}, which comes from a hydrodynamical model that assimilates satellite observations of tidal sea surface heights. 
We first compute
\begin{equation}
    D^2=\frac12(A_{\mathrm{TPXO}}^2-A_{\mathrm{MOM6}}^2)-A_{\mathrm{TPXO}}A_{\mathrm{MOM6}}\cos(\varphi_{\mathrm{TPXO}}-\varphi_{\mathrm{MOM6}}),
\end{equation}
at each grid point, where $A$ is the amplitude of the tidal constituent and $\varphi$ is the Greenwich phase lag. The root mean square error (RMSE) is then defined as
\begin{equation}
    \mathrm{RMSE}=\left(\frac{1}{4\pi R_E^2}\int_SD^2dS\right)^{1/2}.
    \label{eq:RMSE}
\end{equation}
The RMSE can be decomposed into an amplitude error and a phase error~\citep{shriver2012evaluation}. The RMSE is also computed for two different domains: over the entire ocean, and over the deep non-polar ocean, restricting to grid points lying between $66^{\circ}$ south and $66^{\circ}$ north that have a depth of greater than $1000$ meters. The latter computation removes coastal regions and polar regions, where data-assimilative tide models such as TPXO are less accurate~\citep{stammer2014accuracy}. 

\subsection{Tuning}

The parameter $\chi$ in the wave drag scheme is tuned for each method of computing SAL in order to minimize the RMSE. 
One can understand the need for tuning using energy arguments \citep{arbic2004accuracy}; 
if $\chi$ is too large, then too much energy is dissipated, and conversely if $\chi$ is too small, then not enough energy is dissipated. 
Both extremes adversely affect the accuracy of the modeled tides. 
The different methods of computing SAL yield accelerations of different amplitude
and $\chi$ must change accordingly. The $\chi$ parameters used here are presented in Table~\ref{tbl:sal} and Table~\ref{tbl:sal2}. 
Because of differences in how the wave drag was precomputed, the coefficient $\chi$ 
has a different order of magnitude
at $0.36^{\circ}$ and $0.08^{\circ}$ grid spacing.  

\section{Results}
\label{sec:results}

\subsection{TPXO9 Comparison}

\begin{table}[b]
\begin{scalebox}{0.89}{
    \begin{tabular}{lr|rrr|rrr|}
\cline{3-8}\noalign{\vskip-0.45em}
 &
  \multicolumn{1}{l|}{} & 
  \multicolumn{3}{c|}{Deep non-polar ocean} &
  \multicolumn{3}{c|}{Global ocean} \\ \noalign{\vskip-0.2em}\hline 
\multicolumn{1}{|r|}{$0.36^{\circ}$ nominal grid spacing} &
  \multicolumn{1}{l|}{Wave drag $\chi$} &
  \multicolumn{1}{l|}{RMSE} &
  \multicolumn{1}{l|}{Amplitude error} &
  \multicolumn{1}{l|}{Phase error} &
  \multicolumn{1}{l|}{RMSE} &
  \multicolumn{1}{l|}{Amplitude error} &
  \multicolumn{1}{l|}{Phase error} \\ \hline
\multicolumn{1}{|r|}{SHT SAL, $N=40$} &
  0.8 &
  \multicolumn{1}{r|}{6.50} &
  \multicolumn{1}{r|}{3.18} &
  5.70 &
  \multicolumn{1}{r|}{8.65} &
  \multicolumn{1}{r|}{4.63} &
  7.31 \\ \hline
\multicolumn{1}{|r|}{SHT SAL, $N=200$} &
  0.8 &
  \multicolumn{1}{r|}{6.45} &
  \multicolumn{1}{r|}{3.09} &
  5.66 &
  \multicolumn{1}{r|}{8.54} &
  \multicolumn{1}{r|}{4.61} &
  7.18 \\ 
  \hline \multicolumn{1}{|r|}{Ces\`aro SHT SAL, $N=40$} & 0.8 & \multicolumn{1}{r|}{8.10} & \multicolumn{1}{r|}{3.56} & 7.27 & \multicolumn{1}{r|}{10.26} & \multicolumn{1}{r|}{5.01} & 8.95 \\ 
  \hline \multicolumn{1}{|r|}{Ces\`aro SHT SAL, $N=200$} & 0.8 & \multicolumn{1}{r|}{6.76} & \multicolumn{1}{r|}{3.17} & 5.98 & \multicolumn{1}{r|}{8.86} & \multicolumn{1}{r|}{4.68} & 7.52 \\ 
  \hline \multicolumn{1}{|r|}{Ces\`aro SHT SAL, $N=400$} & 0.8 & \multicolumn{1}{r|}{6.60} & \multicolumn{1}{r|}{3.13} & 5.81 & \multicolumn{1}{r|}{8.69} & \multicolumn{1}{r|}{4.65} & 7.35 \\ 
  \hline
\multicolumn{1}{|r|}{Closed Form Convolution SAL} &
  0.7 &
  \multicolumn{1}{r|}{4.10} &
  \multicolumn{1}{r|}{2.78} &
  3.01 &
  \multicolumn{1}{r|}{6.53} &
  \multicolumn{1}{r|}{4.43} &
  4.80 \\ \hline \multicolumn{1}{|r|}{Ces\`aro Tabular Convolution SAL} &
  0.7 &
  \multicolumn{1}{r|}{4.05} &
  \multicolumn{1}{r|}{2.78} &
  2.94 &
  \multicolumn{1}{r|}{6.47} &
  \multicolumn{1}{r|}{4.38} &
  4.77 \\ \hline
\end{tabular}}
\end{scalebox}
\vspace{0.5em}
\caption{Wave drag tuning parameter $\chi$ and errors for different methods of computing SAL with $0.36^{\circ}$ nominal grid spacing.}
\label{tbl:sal}
\end{table}

\begin{table}[h]
\begin{scalebox}{0.89}{
    \begin{tabular}{lr|rrr|rrr|}
\cline{3-8}\noalign{\vskip-0.45em}
 &
  \multicolumn{1}{l|}{} & 
  \multicolumn{3}{c|}{Deep non-polar ocean} &
  \multicolumn{3}{c|}{Global ocean} \\ \noalign{\vskip-0.2em}\hline 
\multicolumn{1}{|r|}{$0.08^{\circ}$ nominal grid spacing} &
  \multicolumn{1}{l|}{Wave drag $\chi$} &
  \multicolumn{1}{l|}{RMSE} &
  \multicolumn{1}{l|}{Amplitude error} &
  \multicolumn{1}{l|}{Phase error} &
  \multicolumn{1}{l|}{RMSE} &
  \multicolumn{1}{l|}{Amplitude error} &
  \multicolumn{1}{l|}{Phase error} \\ \hline
\multicolumn{1}{|r|}{SHT SAL, $N=40$} &
  8.0 &
  \multicolumn{1}{r|}{2.71} &
  \multicolumn{1}{r|}{1.59} &
  2.20 &
  \multicolumn{1}{r|}{5.32} &
  \multicolumn{1}{r|}{3.54} &
  3.96 \\ \hline
\multicolumn{1}{|r|}{SHT SAL, $N=200$} &
  8.0 &
  \multicolumn{1}{r|}{2.67} &
  \multicolumn{1}{r|}{1.55} &
  2.17 &
  \multicolumn{1}{r|}{5.22} &
  \multicolumn{1}{r|}{3.53} &
  3.84 \\ 
  \hline \multicolumn{1}{|r|}{Ces\`aro SHT SAL, $N=40$} & 8.0 & \multicolumn{1}{r|}{4.50} & \multicolumn{1}{r|}{2.34} & 3.85 & \multicolumn{1}{r|}{6.78} & \multicolumn{1}{r|}{3.98} & 5.49 \\ 
  \hline \multicolumn{1}{|r|}{Ces\`aro SHT SAL, $N=200$} & 8.0 & \multicolumn{1}{r|}{3.01} & \multicolumn{1}{r|}{1.68} & 2.49 & \multicolumn{1}{r|}{5.48} & \multicolumn{1}{r|}{3.61} & 4.13 \\ 
  \hline \multicolumn{1}{|r|}{Ces\`aro SHT SAL, $N=400$} & 8.0 & \multicolumn{1}{r|}{2.84} & \multicolumn{1}{r|}{1.61} & 2.33 & \multicolumn{1}{r|}{5.35} & \multicolumn{1}{r|}{3.57} & 3.98 \\ 
  \hline
\multicolumn{1}{|r|}{Closed Form Convolution SAL} &
  8.1 &
   \multicolumn{1}{r|}{2.36} &
  \multicolumn{1}{r|}{1.70} &
  1.63 &
  \multicolumn{1}{r|}{5.06} &
  \multicolumn{1}{r|}{3.64} &
  3.52 \\ \hline \multicolumn{1}{|r|}{Ces\`aro Tabular Convolution SAL} &
  8.1 &
   \multicolumn{1}{r|}{2.32} &
  \multicolumn{1}{r|}{1.69} &
  1.59 &
  \multicolumn{1}{r|}{5.03} &
  \multicolumn{1}{r|}{3.61} & 3.50 \\ \hline
\end{tabular}}
\end{scalebox}
\vspace{0.5em}
\caption{As in Table~\ref{tbl:sal}, but with $0.08^{\circ}$ nominal grid spacing.}
\label{tbl:sal2}
\end{table}

Error results are presented in 
Table~\ref{tbl:sal} with $0.36^{\circ}$ nominal grid spacing
and in
Table~\ref{tbl:sal2} with $0.08^{\circ}$ nominal grid spacing, comparing
the SHT, the Ces\`aro SHT, and the two convolution approaches. 
With $0.36^{\circ}$ nominal grid spacing, 
the $N=40$ SHT has a deep non-polar RMSE of $6.50\;\unit{cm}$, 
with the phase error contributing $5.70\;\unit{cm}$. 
The global RMSE of $8.65\;\unit{cm}$ is similarly dominated by the phase error of $7.31\;\unit{cm}$. 
Using $N=200$ yields only an error reduction of less than $1\%$. 
The Ces\`aro SHT errors decrease from 
$N=40$ to $N=200$ and $N=400$,
yet even for $N=400$ they remain larger than the 
$N=40$ SHT errors.
The convolution errors are smaller than the 
SHT errors in all cases,
with the greatest reduction coming from the phase error. We note that the Ces\`aro tabular convolution and the closed form convolution have similar RMSEs, with the Ces\`aro tabular convolution featuring about a $0.05\;\unit{cm}$ improvement. 
With $0.08^{\circ}$ nominal grid spacing,
significant error reduction is achieved in all cases.
The $N=200$ SHT errors are the smallest among the
five SHT test cases.
The amplitude errors are slightly larger 
with the two convolutions than the $N=200$ SHT,
but this is compensated by smaller phase errors,
so that the convolutions still have the smallest RMSE. As with the error results at $0.36^{\circ}$ nominal grid spacing, the improvement that the Ces\`aro tabular convolution presents over the closed form convolution is small. 

\subsection{Discussion}

The question arises--why are the SHT errors only slightly smaller 
for $N=200$ than for $N=40$?
Figure~\ref{fig:salzonalaccels} and Fig.~\ref{fig:salmeridaccels} show that 
the $N=40$ SHT results have 
small amplitude oscillations in the open ocean, 
while the $N=200$ SHT results have 
similar structure but without the oscillations.
The more significant difference is seen in
Fig.~\ref{fig:sallocalzonalaccels} and Fig.~\ref{fig:sallocalmeridaccels},
where $N=200$ yields larger coastal accelerations 
than $N=40$.
Some of these large coastal accelerations are spurious overshoots due to Gibbs phenomenon, 
aliasing error, and finite differencing error. 
Similar considerations apply to the 
$N=200$ and $N=400$ Ces\`aro SHT results.
Ces\`aro summation might mitigate the impact of Gibbs phenomenon, but it does nothing to remedy errors 
due to aliasing and finite differencing, and as a result, it also results in large coastal accelerations. Additionally, remedying Gibbs errors on the sphere is a subtle matter and Ces\`aro summation may not be sufficient~\citep{gelb1997resolution}. 
By contrast
as seen in Fig.~\ref{fig:sallocalzonalaccels} and Fig.~\ref{fig:sallocalmeridaccels},  
both convolutions predict smaller 
coastal accelerations,
which we attribute to the absence of 
Gibbs phenomenon, aliasing error, and finite differencing error. 

Another question arises--why does the removal
of Gibbs phenomenon and aliasing error in coastal regions reduce the error in the deep non-polar ocean?
We speculate that the global improvement in the tides is due to the back effect of large resonant coastal tides upon open-ocean tides \citep{arbic2007resonance,arbic2009tidal,arbic2010coupled}. 
Thus it appears that improvements to the SAL in coastal regions as a result of using the convolution can drive improvements to the tides in the entire ocean. Further exploration into the role of the back effect in these calculations, 
as well as other potential reasons for the improvement, is left as a topic for future work.  

\subsection{Parallel Efficiency and Runtime}

\begin{figure}[t]
\centering
\includegraphics[width=0.6\linewidth]{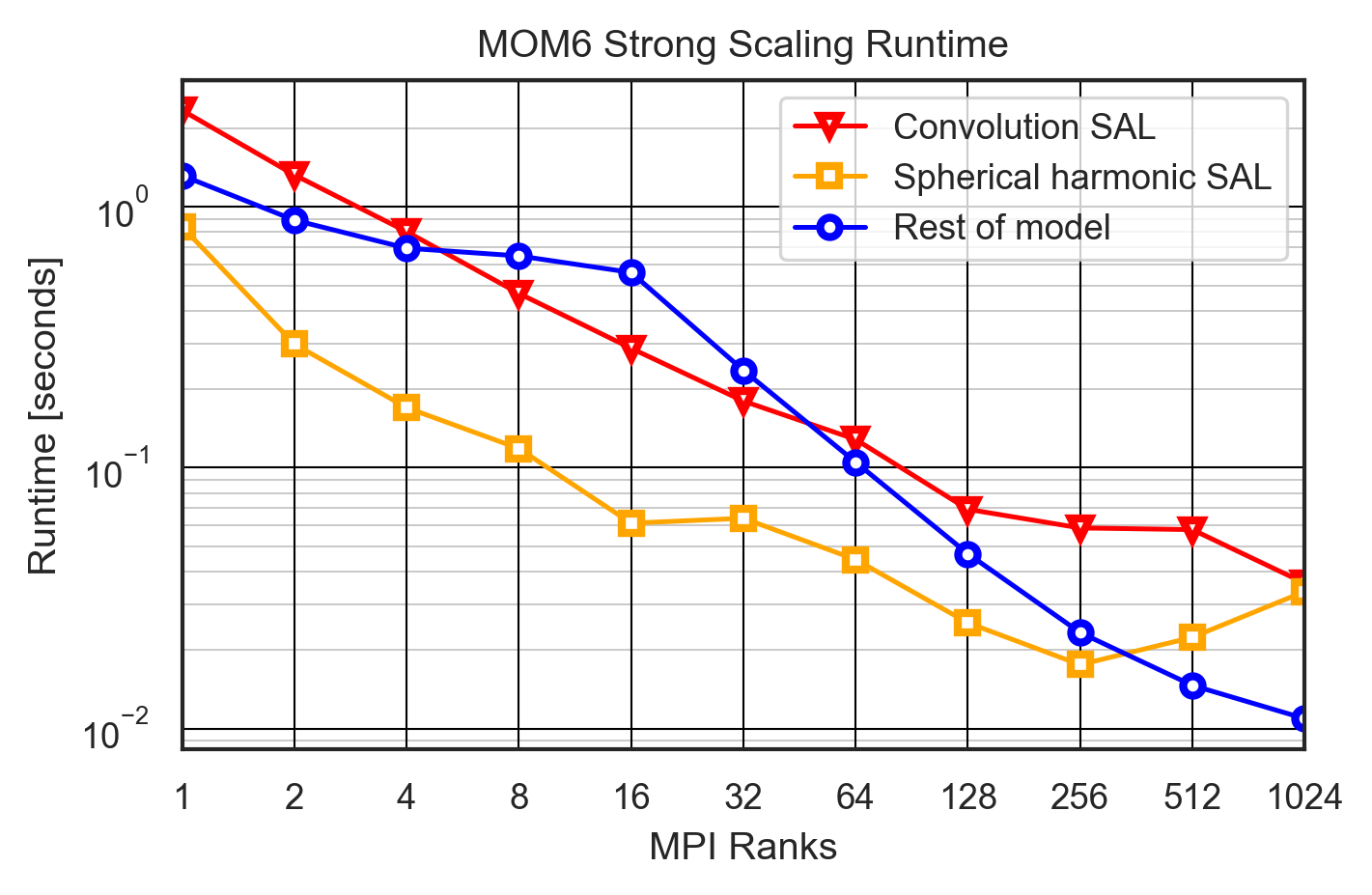}
\caption{Single time step runtime (seconds)
versus number of MPI ranks,
MOM6 running with $0.36^{\circ}$ nominal grid spacing, yielding $N_0=784\,000$ grid cells,
results shown for convolution SAL,
$N=40$ SHT SAL,
non-SAL portion of the calculation.}
\label{fig:runtime}
\end{figure}

A strong scaling analysis is performed to examine
the parallel efficiency and runtime of the convolution/CSFMM SAL.
Figure~\ref{fig:runtime}
shows the single time step runtime (seconds)
versus the number of MPI ranks for MOM6 running 
with $0.36^{\circ}$ nominal grid spacing
yielding $N_0=784\,000$ grid cells.
The runtime is shown for
the convolution/CSFMM SAL,
$N=40$ spherical harmonics SAL, and
non-SAL portion of the calculation.
The data was collected using the
internal performance analysis tools in MOM6.
The SHT and convolution/CSFMM are different
algorithms with different operation counts
and communication patterns,
and this affects their parallel efficiency.
From 1 to 16 ranks,
the SHT scales better and
is faster than the convolution/CSFMM,
but from 16 to 1024 ranks the convolution/CSFMM 
scales better than the SHT.
The SHT runtime increases at 512 and 1024 ranks,
presumably due to communication costs,
while the convolution/CSFMM continues scaling well.
In this single layer simulation
the convolution/CSFMM runtime is comparable to
or smaller than the non-SAL portion 
of the calculation, 
and
its impact in a multi-layer simulation
will be much smaller. The closed form convolution is slightly faster than the Ces\`aro tabular convolution. 

\conclusions 
\label{sec:conclusions}

In the numerical modeling of ocean tides, 
capturing the effects of SAL is necessary 
for accurate prediction of tidal amplitude and phase.
Many current models employ 
spherical harmonics to compute the SAL acceleration. 
This work presents an alternative
where the SAL acceleration is computed by
convolving the sea surface height with
two Green's functions, one based on an asymptotic approximation of the Load Love Numbers, and another based on a precomputed table of values.
The convolution integral is discretized 
on a set of grid cells covering the sphere
and
the sums are accelerated by the
Cubed Sphere Fast Multipole Method. 
The convolution/CSFMM was implemented in the MOM6 ocean model
and
compared with the default SHT
as well as a Ces\`aro modified SHT. 
Errors in tidal amplitude and phase 
were tabulated
using the TPXO9 tidal atlas as the reference.
In these tests the convolution approaches 
have smaller RMSE than the SHT.
The larger SHT error
is attributed to overshoots in
coastal amplitude due to Gibbs phenomenon, 
aliasing error due to finite grid resolution, and finite difference error in computing the
gradient of the SAL potential.
In contrast to the global nature of 
spherical harmonics,
the convolution/CSFMM uses 
local grid cell approximations
which do not exhibit these errors. 
Parallel scaling results up to 1024 MPI ranks show
that the runtime of the convolution/CSFMM
is not prohibitive. 

This work focused on global ocean models,
but incorporating tides and SAL in regional models is also of interest \citep{irazoqui2017effects}. 
Spherical harmonics are poorly suited for that case,
and
in future work we plan to explore the 
convolution/CSFMM SAL in regional models. 
Similarly,
the present work employed a simple discretization 
of the convolution integral~Eq.\eqref{eq:N-body-sum}
and
more accurate schemes
could be considered \citep{agnew2012spotl}.
Finally,
an integral equation method
could bypass the load Love numbers and directly
compute the SAL Green's function on a grid
given a model of the Earth interior.
Such an approach could possibly also incorporate
the inhomogeneity of the Earth interior  \citep{pagiatakis1990response,scherneck1991parametrized}.

\codedataavailability{The current version of MOM6 is available from https://github.com/noaa-gfdl/MOM6 under the GNU Lesser General Public License. The exact version of the model used to produce the results used in this paper is archived on repository under DOI 10.5281/zenodo.18445619 \citep{chen_2026_18445619}, as are input data and scripts to run the model and produce the plots for all the simulations presented in this paper.} 

\authorcontribution{BKA and RK conceptualized the project. AC conducted the formal analysis and investigation. AC and HW produced the software. HW provided datasets. BKA and RK supervised the project. AC visualized the results and prepared the original draft. AC, HW, BKA, and RK assisted in reviewing and editing the manuscript.} 

\competinginterests{The authors declare that they have no conflict of interest.} 

\begin{acknowledgements}

We thank Richard Ray and an anonymous reviewer for useful comments, which greatly improved the content and presentation of the final manuscript. AC was supported by the National Science Foundation Graduate Research Fellowship Program under Grant Number DGE 2241144. Any opinions, findings, and conclusions or recommendations expressed in this material are those of the authors and do not necessarily reflect the views of the National Science Foundation. BKA is grateful for sabbatical support from the Australian National University, especially Callum Shakespeare, Andy Hogg, and Adele Morrison, during the 2024–2025 academic year. BKA acknowledges support from Office of Naval Research grant N00017-22-1-2576. The authors thank Robert Hallberg (National Oceanic and Atmospheric Administration, Geophysical Fluid Dynamics Laboratory) and Alan Wallcraft (Florida State University) for useful discussions. The authors also thank Oliver Jahn (MIT, Program in Atmospheres, Oceans, and Climate) for noticing mistakes in the original statements of Equations 15 and 16, which have since been rectified. We would like to acknowledge that computational work involved in this research work is partially/fully supported by NUS IT’s Research Computing group under grant number NUSREC-HPC-00001.

\end{acknowledgements}

\bibliographystyle{copernicus}
\bibliography{refs.bib}

@article{egbert2004,
  title={Numerical modeling of the global semidiurnal tide in the present day and in the last glacial maximum},
  author={Egbert, Gary D. and Ray, Richard D. and Bills, Bruce G.},
  journal={J. Geophys. Res.},
  volume={109},
  pages={C03003},
  year={2004},
  publisher={Wiley Online Library}
}

@article{guo2004green,
  title={{Green's function of the deformation of the Earth as a result of atmospheric loading}},
  author={Guo, JY and Li, YB and Huang, Y and Deng, HT and Xu, SQ and Ning, JS},
  journal={Geophy. J. Int.},
  volume={159},
  pages={53--68},
  year={2004}
}

@article{hendershott1972effects,
  title={The effects of solid {Earth} deformation on global ocean tides},
  author={Hendershott, MC},
  journal={Geophys. J. Int.},
  volume={29},
  number={4},
  pages={389--402},
  year={1972},
  publisher={Blackwell Publishing Ltd Oxford, UK}
}

@article{adcroft2019gfdl,
  title={The {GFDL} global ocean and sea ice model {OM4}. 0: Model description and simulation features},
  author={Adcroft, Alistair and Anderson, Whit and Balaji, V and Blanton, Chris and Bushuk, Mitchell and Dufour, Carolina O and Dunne, John P and Griffies, Stephen M and Hallberg, Robert and Harrison, Matthew J and others},
  journal={J. Adv. Model. Earth Syst.},
  volume={11},
  number={10},
  pages={3167--3211},
  year={2019},
  publisher={Wiley Online Library}
}

@article{accad1978solution,
  title={Solution of the tidal equations for the {M2} and {S2} tides in the world oceans from a knowledge of the tidal potential alone},
  author={Accad, Y and Pekeris, Chaim L},
  journal={Philos. Trans. R. Soc. London, Ser. A},
  volume={290},
  number={1368},
  pages={235--266},
  year={1978},
  publisher={The Royal Society London}
}

@article{ray1998ocean,
  title={Ocean self-attraction and loading in numerical tidal models},
  author={Ray, Richard D},
  journal={Mar. Geod.},
  volume={21},
  number={3},
  pages={181--192},
  year={1998},
  publisher={Taylor \& Francis}
}

@article{barnes1986hierarchical,
  title={{A hierarchical O(NlogN) force-calculation algorithm}},
  author={Barnes, Josh and Hut, Piet},
  journal={Nature},
  volume={324},
  pages={446--449},
  year={1986}
}

@article{greengard1987fast,
  title={A fast algorithm for particle simulations},
  author={Greengard, Leslie and Rokhlin, Vladimir},
  journal={J. Comput. Phys.},
  volume={73},
  pages={325--348},
  year={1987}
}

@article{berrut2004barycentric,
  title={Barycentric Lagrange interpolation},
  author={Berrut, Jean-Paul and Trefethen, Lloyd N},
  journal={SIAM Review},
  volume={46},
  pages={501--517},
  year={2004}
}

@article{arbic2004accuracy,
  title={The accuracy of surface elevations in forward global barotropic and baroclinic tide models},
  author={Arbic, Brian K and Garner, Stephen T and Hallberg, Robert W and Simmons, Harper L},
  journal={Deep Sea Res. Part II},
  volume={51},
  number={25-26},
  pages={3069--3101},
  year={2004},
  publisher={Elsevier}
}

@article{barton2022global,
  title={Global barotropic tide modeling using inline self-attraction and loading in {MPAS-Ocean}},
  author={Barton, Kristin N and Pal, Nairita and Brus, Steven R and Petersen, Mark R and Arbic, Brian K and Engwirda, Darren and Roberts, Andrew F and Westerink, Joannes J and Wirasaet, Damrongsak and Schindelegger, Michael},
  journal={J. Adv. Model. Earth Syst.},
  volume={14},
  number={11},
  pages={e2022MS003207},
  year={2022},
  publisher={Wiley Online Library}
}

@article{brus2023scalable,
  title={Scalable self attraction and loading calculations for unstructured ocean tide models},
  author={Brus, Steven R and Barton, Kristin N and Pal, Nairita and Roberts, Andrew F and Engwirda, Darren and Petersen, Mark R and Arbic, Brian K and Wirasaet, Damrongsak and Westerink, Joannes J and Schindelegger, Michael},
  journal={Ocean Modell.},
  volume={182},
  pages={102160},
  year={2023},
  publisher={Elsevier}
}

@article{farrell1972deformation,
  title={Deformation of the {Earth} by surface loads},
  author={Farrell, W E},
  journal={Rev. Geophys.},
  volume={10},
  number={3},
  pages={761--797},
  year={1972},
  publisher={Wiley Online Library}
}

@article{wang2012load,
  title={Load {Love} numbers and {Green}'s functions for elastic {Earth} models {PREM, iasp91, ak135}, and modified models with refined crustal structure from {Crust} 2.0},
  author={Wang, Hansheng and Xiang, Longwei and Jia, Lulu and Jiang, Liming and Wang, Zhiyong and Hu, Bo and Gao, Peng},
  journal={Comput. Geosci.},
  volume={49},
  pages={190--199},
  year={2012},
  publisher={Elsevier}
}

@article{jayne2001parameterizing,
  title={Parameterizing tidal dissipation over rough topography},
  author={Jayne, Steven R and St. Laurent, Louis C},
  journal={Geophys. Res. Lett.},
  volume={28},
  number={5},
  pages={811--814},
  year={2001},
  publisher={Wiley Online Library}
}

@article{buijsman2015optimizing,
  title={Optimizing internal wave drag in a forward barotropic model with semidiurnal tides},
  author={Buijsman, Maarten C and Arbic, Brian K and Green, JAM and Helber, Robert W and Richman, James G and Shriver, Jay F and Timko, PG and Wallcraft, A},
  journal={Ocean Modell.},
  volume={85},
  pages={42--55},
  year={2015},
  publisher={Elsevier}
}

@book{gradshteyn2014table,
  title={{Table of integrals, series, and products}},
  author={Gradshteyn, Izrail Solomonovich and Ryzhik, Iosif Moiseevich},
  year={2014},
  publisher={Academic press}
}

@inproceedings{erofeeva2018tpxo9,
  title={TPXO9-a new global tidal model in TPXO series},
  author={Erofeeva, Svetlana and Egbert, Gary D},
  booktitle={American Geophysical Union, Ocean Sciences Meeting},
  number={186},
  pages={PL34C--186},
  year={2018}
}

@article{chen2025cubed,
  title={A Cubed Sphere Fast Multipole Method},
  author={Chen, Anthony and Krasny, Robert},
  journal={arXiv preprint arXiv:2508.13550},
  year={2025}
}

@article{wang2024improving,
  title={Improving global barotropic tides with sub-grid scale topography},
  author={Wang, He and Hallberg, Robert and Wallcraft, Alan J and Arbic, Brian K and Chassignet, Eric P},
  journal={J. Adv. Model. Earth Syst.},
  volume={16},
  number={4},
  pages={e2023MS004056},
  year={2024},
  publisher={Wiley Online Library}
}

@article{arbic2022incorporating,
  title={Incorporating tides and internal gravity waves within global ocean general circulation models: A review},
  author={Arbic, Brian K},
  journal={Prog. Oceanogr.},
  volume={206},
  pages={102824},
  year={2022},
  publisher={Elsevier}
}

@article{arbic2010concurrent,
  title={Concurrent simulation of the eddying general circulation and tides in a global ocean model},
  author={Arbic, Brian K and Wallcraft, Alan J and Metzger, E Joseph},
  journal={Ocean Modell.},
  volume={32},
  number={3-4},
  pages={175--187},
  year={2010},
  publisher={Elsevier}
}

@article{muller2010effect,
  title={The effect of ocean tides on a climate model simulation},
  author={M{\"u}ller, M and Haak, Helmut and Jungclaus, Johann H and S{\"u}ndermann, J and Thomas, Maik},
  journal={Ocean Modell.},
  volume={35},
  number={4},
  pages={304--313},
  year={2010},
  publisher={Elsevier}
}

@article{le1994spectroscopy,
  title={Spectroscopy of the world ocean tides from a finite element hydrodynamic model},
  author={Le Provost, C and Genco, ML and Lyard, F and Vincent, P and Canceil, P},
  journal={J. Geophys. Res.: Oceans},
  volume={99},
  number={C12},
  pages={24777--24797},
  year={1994},
  publisher={Wiley Online Library}
}

@article{parke1980m2,
  title={{M2}, {S2}, {K1} models of the global ocean tide on an elastic {Earth}},
  author={Parke, Michael E and Hendershott, Myrl C},
  journal={Mar. Geod.},
  volume={3},
  number={1-4},
  pages={379--408},
  year={1980},
  publisher={Taylor \& Francis}
}

@article{ray2002global,
  title={Global ocean tide models on the eve of {TOPEX/POSEIDON}},
  author={Ray, Richard D},
  journal={IEEE Trans. Geosci. Remote Sens.},
  volume={31},
  number={2},
  pages={355--364},
  year={1993},
  publisher={IEEE}
}

@book{schwiderski1979global,
  title={On charting global ocean tides},
  author={Schwiderski, Ernst W},
  journal={Reviews of geophysics},
  volume={18},
  number={1},
  pages={243--268},
  year={1980},
  publisher={Wiley Online Library}
}

@article{shum1997accuracy,
  title={Accuracy assessment of recent ocean tide models},
  author={Shum, CK and Woodworth, PL and Andersen, OB and Egbert, Gary D and Francis, Olivier and King, C and Klosko, SM and Le Provost, Ch and Li, X and Molines, J-M and others},
  journal={J. Geophys. Res.: Oceans},
  volume={102},
  number={C11},
  pages={25173--25194},
  year={1997},
  publisher={Wiley Online Library}
}

@article{cummins1997simulation,
  title={Simulation of barotropic and baroclinic tides off northern {British Columbia}},
  author={Cummins, Patrick F and Oey, Lie-Yauw},
  journal={J. Phys. Oceanogr.},
  volume={27},
  number={5},
  pages={762--781},
  year={1997}
}

@article{kang2002two,
  title={Two-layer tidal modeling of the {Yellow} and {East China} Seas with application to seasonal variability of the {M2} tide},
  author={Kang, Sok Kuh and Foreman, Michael GG and Lie, Heung-Jae and Lee, Jae-Hak and Cherniawsky, Josef and Yum, Ki-Dai},
  journal={J. Geophys. Res.: Oceans},
  volume={107},
  number={C3},
  pages={6--1},
  year={2002},
  publisher={Wiley Online Library}
}

@article{merrifield2001generation,
  title={The generation of internal tides at the {Hawaiian} Ridge},
  author={Merrifield, Mark A and Holloway, Peter E and Johnston, TM Shaun},
  journal={Geophys. Res. Lett.},
  volume={28},
  number={4},
  pages={559--562},
  year={2001},
  publisher={Wiley Online Library}
}

@article{schiller2007explicit,
  title={Explicit tidal forcing in an ocean general circulation model},
  author={Schiller, A and Fiedler, R},
  journal={Geophys. Res. Lett.},
  volume={34},
  number={3},
  year={2007},
  publisher={Wiley Online Library}
}

@article{thomas2001consideration,
  title={Consideration of ocean tides in an {OGCM} and impacts on subseasonal to decadal polar motion excitation},
  author={Thomas, Maik and S{\"u}ndermann, J{\"u}rgen and Maier-Reimer, Ernst},
  journal={Geophys. Res. Lett.},
  volume={28},
  number={12},
  pages={2457--2460},
  year={2001},
  publisher={Wiley Online Library}
}

@article{gordeev1977effects,
  title={The effects of loading and self-attraction on global ocean tides: The model and the results of a numerical experiment},
  author={Gordeev, RG and Kagan, BA and Polyakov, EV},
  journal={J. Phys. Oceanogr.},
  volume={7},
  number={2},
  pages={161--170},
  year={1977}
}

@article{stepanov2004parameterization,
  title={Parameterization of ocean self-attraction and loading in numerical models of the ocean circulation},
  author={Stepanov, Vladimir N and Hughes, Christopher W},
  journal={J. Geophys. Res.: Oceans},
  volume={109},
  number={C3},
  year={2004},
  publisher={Wiley Online Library}
}

@article{irazoqui2017effects,
  title={Effects of self-attraction and loading at a regional scale: a test case for the {Northwest} {European} Shelf},
  author={Irazoqui Apecechea, Maialen and Verlaan, Martin and Zijl, Firmijn and Le Coz, Camille and Kernkamp, Herman},
  journal={Ocean Dyn.},
  volume={67},
  number={6},
  pages={729--749},
  year={2017},
  publisher={Springer}
}

@article{scherneck1991parametrized,
  title={A parametrized solid {Earth} tide model and ocean tide loading effects for global geodetic baseline measurements},
  author={Scherneck, H-G},
  journal={Geophys. J. Int.},
  volume={106},
  number={3},
  pages={677--694},
  year={1991},
  publisher={Blackwell Publishing Ltd Oxford, UK}
}

@article{pagiatakis1990response,
  title={The response of a realistic {Earth} to ocean tide loading},
  author={Pagiatakis, Spiros D},
  journal={Geophys. J. Int.},
  volume={103},
  number={2},
  pages={541--560},
  year={1990},
  publisher={Blackwell Publishing Ltd Oxford, UK}
}

@Manual{derecho,
  author       = {{Computational and Information Systems Laboratory}},
  organization = {{NSF: National Center for Atmospheric Research }},
  title        = {{Derecho: HPE Cray EX System (University Community Computing)}},
  note         = {doi:10.5065/qx9a-pg09},
  year         = {2023},
  address      = {Boulder, CO}
}

@article{arbic2007resonance,
  title={On the resonance and influence of the tides in {Ungava} Bay and {Hudson} Strait},
  author={Arbic, Brian K and St-Laurent, Pierre and Sutherland, Graig and Garrett, Chris},
  journal={Geophys. Res. Lett.},
  volume={34},
  number={17},
  year={2007},
  publisher={Wiley Online Library}
}

@article{arbic2009tidal,
  title={On tidal resonance in the global ocean and the back-effect of coastal tides upon open-ocean tides},
  author={Arbic, Brian K and Karsten, Richard H and Garrett, Chris},
  journal={Atmos. Ocean},
  volume={47},
  number={4},
  pages={239--266},
  year={2009},
  publisher={Taylor \& Francis}
}

@article{arbic2010coupled,
  title={A coupled oscillator model of shelf and ocean tides},
  author={Arbic, Brian K and Garrett, Chris},
  journal={Cont. Shelf Res.},
  volume={30},
  number={6},
  pages={564--574},
  year={2010},
  publisher={Elsevier}
}

@article{egbert1994topex,
  title={{TOPEX/POSEIDON} tides estimated using a global inverse model},
  author={Egbert, Gary D and Bennett, Andrew F and Foreman, Michael GG},
  journal={J. Geophys. Res.: Oceans},
  volume={99},
  number={C12},
  pages={24821--24852},
  year={1994},
  publisher={Wiley Online Library}
}

@article{munk1998abyssal,
  title={Abyssal recipes II: Energetics of tidal and wind mixing},
  author={Munk, Walter and Wunsch, Carl},
  journal={Deep Sea Res. Part I},
  volume={45},
  number={12},
  pages={1977--2010},
  year={1998},
  publisher={Elsevier}
}

@article{simpson1974fronts,
  title={Fronts in the {Irish} sea},
  author={Simpson, John H and Hunter, JR},
  journal={Nature},
  volume={250},
  number={5465},
  pages={404--406},
  year={1974},
  publisher={Nature Publishing Group UK London}
}

@article{gottlieb1997gibbs,
  title={On the {Gibbs} phenomenon and its resolution},
  author={Gottlieb, David and Shu, Chi-Wang},
  journal={SIAM Rev.},
  volume={39},
  number={4},
  pages={644--668},
  year={1997},
  publisher={SIAM}
}

@article{stammer2014accuracy,
  title={Accuracy assessment of global barotropic ocean tide models},
  author={Stammer, Detlef and Ray, Richard D and Andersen, Ole Baltazar and Arbic, Brian K and Bosch, Wolfgang and Carrere, Laurent and Cheng, Yongcun and Chinn, Douglas S and Dushaw, Brian D and Egbert, Gary D and others},
  journal={Rev. Geophys.},
  volume={52},
  number={3},
  pages={243--282},
  year={2014},
  publisher={Wiley Online Library}
}

@article{waterhouse2014global,
  title={Global patterns of diapycnal mixing from measurements of the turbulent dissipation rate},
  author={Waterhouse, Amy F and MacKinnon, Jennifer A and Nash, Jonathan D and Alford, Matthew H and Kunze, Eric and Simmons, Harper L and Polzin, Kurt L and St. Laurent, Louis C and Sun, Oliver M and Pinkel, Robert and others},
  journal={J. Phys. Oceanogr.},
  volume={44},
  number={7},
  pages={1854--1872},
  year={2014}
}

@software{chen_2026_18445619,
  author       = {Chen, Anthony and
                  Wang, He},
  title        = {Convolution Based Self Attraction and Loading in
                   MOM6
                  },
  month        = jan,
  year         = 2026,
  publisher    = {Zenodo},
  doi          = {10.5281/zenodo.18445619},
  url          = {https://doi.org/10.5281/zenodo.18445619},
}

@article{farrell1973discussion,
  title={A discussion on the measurement and interpretation of changes of strain in the Earth-Earth tides, ocean tides and tidal loading},
  author={Farrell, W E},
  journal={Philosophical Transactions of the Royal Society of London. Series A, Mathematical and Physical Sciences},
  volume={274},
  number={1239},
  pages={253--259},
  year={1973},
  publisher={The Royal Society London}
}

@article{schrama2005three,
  title={Three algorithms for the computation of tidal loading and their numerical accuracy},
  author={Schrama, Ernst J. O.},
  journal={Journal of Geodesy},
  volume={78},
  number={11},
  pages={707--714},
  year={2005},
  publisher={Springer}
}

@article{martens2019loaddef,
  title={LoadDef: A Python-based toolkit to model elastic deformation caused by surface mass loading on spherically symmetric bodies},
  author={Martens, Hilary R and Rivera, Luis and Simons, Mark},
  journal={Earth and Space Science},
  volume={6},
  number={2},
  pages={311--323},
  year={2019},
  publisher={Wiley Online Library}
}

@article{agnew2012spotl,
  title={SPOTL: Some programs for ocean-tide loading},
  author={Agnew, Duncan Carr},
  year={2012}
}

@article{merriam1980series,
  title={The series computation of the gravitational perturbation due to an ocean tide},
  author={Merriam, JB},
  journal={Physics of the Earth and Planetary Interiors},
  volume={23},
  number={2},
  pages={81--86},
  year={1980},
  publisher={Elsevier}
}

@article{shriver2012evaluation,
  title={An evaluation of the barotropic and internal tides in a high-resolution global ocean circulation model},
  author={Shriver, JF and Arbic, Brian K and Richman, JG and Ray, Richard D and Metzger, EJ and Wallcraft, AJ and Timko, PG},
  journal={Journal of Geophysical Research: Oceans},
  volume={117},
  number={C10},
  year={2012},
  publisher={Wiley Online Library}
}

@article{jekeli1996spherical,
  title={Spherical harmonic analysis, aliasing, and filtering},
  author={Jekeli, Christopher},
  journal={Journal of Geodesy},
  volume={70},
  number={4},
  pages={214--223},
  year={1996},
  publisher={Springer}
}

@article{sanso1990aliasing,
  title={On the aliasing problem in the spherical harmonic analysis},
  author={Sans{\`o}, F},
  journal={Bulletin g{\'e}od{\'e}sique (1976-1995)},
  volume={64},
  number={4},
  pages={313--330},
  year={1990},
  publisher={Springer}
}

@article{rocha2016mesoscale,
  title={Mesoscale to submesoscale wavenumber spectra in Drake Passage},
  author={Rocha, Cesar B and Chereskin, Teresa K and Gille, Sarah T and Menemenlis, Dimitris},
  journal={Journal of Physical Oceanography},
  volume={46},
  number={2},
  pages={601--620},
  year={2016}
}

@article{schindelegger2018can,
  title={Can we model the effect of observed sea level rise on tides?},
  author={Schindelegger, Michael and Green, JAM and Wilmes, S-B and Haigh, Ivan David},
  journal={Journal of Geophysical Research: Oceans},
  volume={123},
  number={7},
  pages={4593--4609},
  year={2018},
  publisher={Wiley Online Library}
}

@article{desai2014consideration,
  title={Consideration of tidal variations in the geocenter on satellite altimeter observations of ocean tides},
  author={Desai, Shailen D and Ray, Richard D},
  journal={Geophysical Research Letters},
  volume={41},
  number={7},
  pages={2454--2459},
  year={2014},
  publisher={Wiley Online Library}
}

@article{weyl1910gibbs,
  title={Die gibbs’ sche erscheinung in der theorie der kugelfunktionen},
  author={Weyl, Hermann},
  journal={Rendiconti del Circolo Matematico di Palermo (1884-1940)},
  volume={29},
  number={1},
  pages={308--323},
  year={1910},
  publisher={Springer}
}

@article{gelb1997resolution,
  title={The resolution of the Gibbs phenomenon for spherical harmonics},
  author={Gelb, Anne},
  journal={Math. Comput.},
  volume={66},
  number={218},
  pages={699--717},
  year={1997}
}

\end{document}